\documentclass[prd,twocolumn,showpacs]{revtex4-1}
\usepackage{mathrsfs}
\usepackage{amsmath}
\usepackage{amssymb}
\usepackage{bm}
\usepackage{amsfonts}
\usepackage{subfigure}
\usepackage{feynmp}
\usepackage{extarrows}
\usepackage{slashed}
\usepackage{graphicx}

\usepackage[linktocpage=true,bookmarks=true,bookmarksnumbered=true,breaklinks=true,pdfpagemode=Fullscreen,pdfstartview=FitBH]{hyperref}

\renewcommand{\FR}[2]{\displaystyle\frac{\,{#1}\,}{#2}}
\newcommand{\fr}[2]{\mbox{$\frac{\,{#1}\,}{#2}$}}
\newcommand{\n}{\nonumber}

\def\bge{\begin{equation}}
\def\ede{\end{equation}}
\def\bga{\begin{aligned}}
\def\eda{\end{aligned}}
\newcommand{\beq}{\begin{equation}}
\newcommand{\eeq}{\end{equation}}
\newcommand{\bq}{\begin{equation}}
\newcommand{\eq}{\end{equation}}
\newcommand{\ba}{\begin{array}}
\newcommand{\ea}{\end{array}}
\newcommand{\beqa}{\begin{eqnarray}}
\newcommand{\eeqa}{\end{eqnarray}}
\newcommand{\beqs}{\begin{subequations}}
\newcommand{\eeqs}{\end{subequations}}

\def\dis{\displaystyle}
\def\ff{\frac}
\def\({\left(}
\def\){\right)}

\def\hf{\frac{1}{2}}

\def\End{\end{document}}
\def\leqq{\leqslant}

\def\hf{\frac{1}{2}}

\newcommand{\order}[1]{\mathcal{O}({#1})}
\def\di{{\mathrm{d}}}
\def\D{{\mathrm{D}}}

\def\T{\mathcal{T}}

\def\pd{\partial}
\def\ld{{\mathcal{L}}}

\def\ra{\rangle}
\def\sla{\slashed}

\def\to{\rightarrow}

\def\ii{\mathrm{i}}

\def\ga{\gamma}
\def\de{\delta}

\def\lam{\lambda}
\def\rh{\rho}
\def\si{\sigma}

\def\Mp{M_{\mathrm{Pl}}}

\def\cut{\Lambda_{\text{UV}}^{}}

\begin{document}

\title{Gravitational Interaction of Higgs Boson and Weak Boson Scattering}

\author{{\sc Zhong-Zhi Xianyu} and {\sc Jing Ren}}
\affiliation{Institute of Modern Physics and Center for High Energy Physics,
Tsinghua University, Beijing 100084, China}
\author{{\sc Hong-Jian He}\,}
\email[Corresponding author ]{(hjhe@tsinghua.edu.cn)}
\affiliation{
Institute of Modern Physics and Center for High Energy Physics,
Tsinghua University, Beijing 100084, China
\\
Center for High Energy Physics, Peking University, Beijing 100871, China
\\
Kavli Institute for Theoretical Physics China, CAS, Beijing 100190, China
}


\begin{abstract}
With the LHC discovery of a 125\,GeV Higgs-like boson, we study gravitational interaction of
Higgs boson via the unique dimension-4 operator involving Higgs doublet and scalar curvature,
$\,\xi H^\dag\! H \mathcal{R}\,$,\, with nonminimal coupling $\,\xi\,$.\,
This Higgs portal term can be transformed away in Einstein frame and
induces gauge-invariant effective interactions in the Higgs sector.
We study the weak boson scattering in Einstein frame, and explicitly demonstrate the
longitudinal-Goldstone boson equivalence theorem in the presence of $\,\xi\,$ coupling.
With these, we derive perturbative unitarity bound on the Higgs gravitational coupling $\,\xi\,$
in Einstein frame, which is stronger than that inferred from the LHC
Higgs measurements.  We further study $\xi$-dependent weak boson scattering cross sections
at TeV scale, and propose a new LHC probe of the Higgs-gravity coupling $\,\xi\,$
via weak boson scattering experiments.
\\[1.5mm]
\hfill Phys.\ Rev.\ D (2013), in Press [arXiv:1305.0251]
\end{abstract}


\maketitle

\noindent
{\bf 1.\,Introduction}
\vspace*{2mm}

The Standard Model (SM) of electromagnetic, weak and strong interactions has been extremely successful.
It has the well established $\,SU(3)_c^{}\otimes SU(2)_L^{}\otimes U(1)_Y^{}\,$ gauge structure,
and hypothesizes a single Higgs doublet to generate masses for weak gauge bosons as well as
all three families of quarks, leptons and neutrinos.
The recent LHC discovery of a 125\,GeV Higgs-like boson \cite{LHC2012-7,LHC2013-3} appears to
provide the last missing piece of the SM. However, the SM is definitely incomplete for many
reasons --- above all, the SM does not contain gravitational force. At present, the best
theory of gravitation is Einstein's general relativity (GR), but is notoriously
nonrenormalizable \cite{HV} despite that the gravity itself is still weakly coupled
over vast energy ranges (below Planck mass) and thus the perturbation should hold well.
Given the fact that all SM particles participate in gravitation, we have to treat
both the SM and GR together as a joint low energy effective theory below Planck scale.

\vspace*{1.5mm}

In the modern effective theory formulation \cite{EFT}, we study low energy physics
by writing down the most general operators in the Lagrangian, under perturbative expansion
and consistent with all given symmetries. Indeed, the SM gives
such a general effective Lagrangian up to dimension-4 operators under the
$\,SU(3)_c^{}\otimes SU(2)_L^{}\otimes U(1)_Y^{}\,$ gauge structure and
the known particle content.
On the other hand, the GR of gravitation is described by Einstein-Hilbert action, containing
two leading terms of a series of operators consistent with general covariance,
\beqa
S_{\text{GR}}^{} \,=\, \Mp^2\int\!\di^4x\,\sqrt{-g\,}\(-\Lambda+\fr{1}{2}\mathcal{R}\),
\hspace*{4mm}
\label{eq:GR0}
\eeqa
where $\,\Lambda\,$ denotes the cosmological constant, $\,\mathcal{R}$\, is the Ricci scalar,
and $\,\Mp=(8\pi G)^{-1/2}\simeq 2.44 \times 10^{18}$\,GeV gives the reduced Planck mass.
To complete this series of operators up to dimension-4, we include two more terms,
\beqa
\label{eq:GR1}
\Delta S_{\text{GR}}^{} \,=\,
\int\!\di^4x\,\sqrt{-g\,}
\(c_1^{} \mathcal{R}^2 + c_2^{}\mathcal{R}_{\mu\nu}^{}\mathcal{R}^{\mu\nu}_{}\).
~~~
\eeqa
The other dimension-4 term $\,\mathcal{R}_{\mu\nu\rh\si}\mathcal{R}^{\mu\nu\rh\si}$\,
is not independent up to integration by parts. 

\vspace*{1.5mm}

Combining the SM and GR together, one usually requires that the SM particles couple to gravity
through their energy-momentum tensor, so the resultant theory is consistent with
the SM gauge symmetries and the equivalence principle.
Practically, this amounts to the replacement
$\,\eta_{\mu\nu}^{}\to g_{\mu\nu}^{}$\, and
$\,\pd_\mu^{}\to \nabla_\mu^{}\,$ in the SM Lagrangian,
where $\,\nabla^{}_\mu\,$ is the covariant derivative associated with metric
$\,g_{\mu\nu}^{}\,$.\, (For fermions, vierbein and spin connection will be used.)
This gives the so-called minimal couplings between the SM and gravity.

\vspace*{1.5mm}

In light of the recent LHC Higgs discovery, we are strongly motivated to study
gravitational interactions of Higgs boson because Higgs boson is the origin
of inertial mass generation for elementary particles.
For this, there is a unique dimension-4 Higgs portal operator
that should be included in the SM\,+\,GR system,
\beqa
\Delta S_{\text{NMC}}^{} \,=\, \xi\int\!\di^4x\,\sqrt{-g\,}\,H^\dag H \mathcal{R} \,,
\hspace*{6mm}
\label{eq:NMC}
\eeqa
where $\,H\,$ denotes the SM Higgs doublet, and the dimensionless parameter $\,\xi\,$
is conventionally called the nonminimal coupling which describes Higgs-curvature interaction.
This dimension-4 operator respects all symmetries of the SM and has to be retained in the effective
theory formulation of the SM+GR. Even if we naively set the nonminimal coupling $\,\xi =0\,$
at tree-level, it will reappear at loop level \cite{RGxi}.
Furthermore, since setting $\,\xi =0\,$ does not
enhance the symmetry, there is no reason that $\,\xi\,$ would be small.
A special choice of $\,\xi=-1/6$\, will make the theory conformally invariant,
but the conformal symmetry is not respected by the SM Lagrangian,
and in most cases will be further broken by quantum effects.
Hence, such a dimensionless coupling $\,\xi\,$ could be rather large, \emph{a priori.}
This fact has already been put in use for the Higgs inflation scenario,
where $\,\xi\gg 1$\, is required, and the typical input is $\,\xi\sim 10^{4}\,$ \cite{bezrukov,ext}.
A recent interesting study by Atkins and Calmet
derived a bound on $\,\xi\,$ from the LHC data,
$\,|\xi|< 2.6\times 10^{15}$\, \cite{atkins}.

\vspace*{1.5mm}

In this work, we study the Higgs gravitational coupling in Einstein frame (Sec.\,2),
where the original operator (\ref{eq:NMC}), as defined in Jordan frame,
is transformed away and induces effective interactions for Higgs sector.
These induced new Higgs interactions modify Higgs couplings
with weak gauge bosons and Goldstone bosons.
With this formulation, we analyze the scattering amplitudes of longitudinal weak bosons and
Goldstone bosons in Sec.\,3.
We give the {\it first demonstration} of longitudinal-Goldstone boson equivalence theorem
with nonzero $\,\xi\,$ coupling. Then, we derive perturbative unitarity bound of $\,\xi\,$
in Einstein frame, to leading order of $\Mp^{-2}$.\,
As we will show, the bound can be significantly stronger than the current Higgs search bound
\cite{atkins}, due to the quadratical dependence of the amplitude on the scattering energy.
Furthermore, we study an intriguing scenario of SM+GR effective theory
with a low ultraviolet (UV) cutoff around $\,\order{10\,\text{TeV}}$,\,
as a natural solution to the hierarchy problem.
With this we make {\it the first proposal} to study the impacts of
$\,\xi\,$ coupling on TeV-scale weak boson scattering
cross sections and discuss the LHC tests.
Our final conclusion and discussion will be given in Sec.\,4.

\vspace*{5mm}
\noindent
{\bf 2.\,Induced Higgs Portal Interactions for Einstein Frame}
\vspace*{3mm}

Even though the SM has been impressively consistent with all the experimental data so far
--- including the recent LHC discovery of a Higgs-like particle \cite{LHC2012-7,LHC2013-3},
the unique dimension-4 Higgs-gravity interaction (\ref{eq:NMC}) is an unambiguous
portal to the new physics beyond SM.
It is possible that the effective theory combining the SM with general relativity
(\ref{eq:GR0})-(\ref{eq:GR1}) via the Higgs portal (\ref{eq:NMC}) will describe the Universe
all the way up to very high scales (below Planck mass $\Mp$),
including inflation as well (with proper extensions) \cite{bezrukov}\cite{ext}.
For the electroweak gauge and Higgs sectors, we can write this effective theory
up to dimension-4 operators, as the following,
\begin{align}
\label{S-Jordan}
\hspace*{-3mm}
S \,=& \int\!\!\di^4x\,\sqrt{-g^{(J)}}
\Big[\!-\Mp^2\Lambda \!+\! \big(\fr{\,M^2}{2}\! + \xi H^\dag H\big)\mathcal{R}^{(J)}~~
\n\\
\hspace*{-2.5mm}
    &
    -\fr{1}{4}F_{j\mu\nu}^a F^{a\mu\nu}_j\!
    +(\D_\mu H)^\dag(\D^\mu H)-V(H)\Big] ,~~
\end{align}
where $\,\mathcal{R}^{(J)}\,$ is the scalar curvature associated with metric
$\,g_{\mu\nu}^{(J)}\,$,\, and $\,F^{a\mu\nu}_j(=W^{a\mu\nu}_{},B^{\mu\nu}_{})$\,
denotes field strengths of
$\,SU(2)_L^{}\otimes U(1)_Y^{}\,$ electroweak gauge fields.
The Higgs doublet $\,H\,$ has the potential,
$\,V(H)=\lambda (H^\dag H-\fr{1}{2}v^2)^2\,$.\,
The frame including the $\xi$-term in (\ref{S-Jordan}) is conventionally called Jordan frame.
Here we have used a superscript ``${(J)}$" for the metric
$\,g_{\mu\nu}^{(J)}\,$ and Ricci scalar $\,\mathcal{R}^{(J)}\,$.\,
One can always make a field-dependent Weyl transformation
for the Jordan frame metric $\,g_{\mu\nu}^{(J)}\,$
such that this $\xi$-term is transformed away,
and the resultant system is called Einstein frame
in which the field equation of $\,g_{\mu\nu}^{}\,$  is the standard
Einstein equation for GR.  The field redefinition in Einstein frame reads,
$\,g_{\mu\nu}^{} = \Omega^2 g_{\mu\nu}^{(J)}$\,,\,
where $\,\Omega^2=(M^2+2\xi H^\dag H)/\Mp^2$\,.\,
For simplicity, hereafter we will denote all quantities in the Einstein frame without extra
superscripts.  From (\ref{S-Jordan}) we can identify the Planck mass,
$\,\Mp^2 = M^2+\xi v^2\,$,\,
where $\,v\,$
is the Higgs vacuum expectation value.
It shows that the nonminimal coupling strength $\,\xi\,$ is bounded by
$\,\xi \leqq \Mp^2/v^2\simeq 9.8 \times 10^{31}$\,,\,
which is saturated in the limit $\,M=0\,$.\,

\vspace*{1.5mm}

Given the above relation between $\,g^{(J)}_{\mu\nu}\,$ and $\,g^{}_{\mu\nu}\,$,
the Ricci scalar transforms as follows,
\begin{align}
  \mathcal{R}^{(J)}
  \,=\,&~ \Omega^2\big[\mathcal{R}-6g^{\mu\nu}_{}\nabla_\mu^{}\nabla_\nu^{}\log\Omega
\n\\
       &~ \hspace*{5mm} +6g^{\mu\nu}_{}(\nabla_\mu^{}\log\Omega)(\nabla_\nu^{}\log\Omega)\big],
\hspace*{10mm}
\end{align}
where the covariant derivative $\,\nabla_\mu^{}\,$ is associated with Einstein frame metric
$\,g_{\mu\nu}^{}$\,.\,
Thus, the effective action (\ref{S-Jordan}) can be rewritten in Einstein frame,
\begin{align}
\label{S-Einstein}
S ~=&~ \int\!\di^4x\,\sqrt{-g\,}\,
\Big[-\ff{\Mp^2}{\Omega^4}\Lambda + \ff{\Mp^2}{2} \mathcal{R}-\ff{1}{4}F_{j\mu\nu}^aF^{a\mu\nu}_j
\hspace*{4mm}
\n\\
    &~~~~-\ff{3\xi}{\Omega^2}\nabla^2(H^\dag H)
         +\ff{9\xi^2}{\Mp^2\Omega^4}\big(\nabla_\mu^{}(H^\dag H)\big)^2
\n\\
    &~~~~+\ff{1}{\Omega^2}(\D^\mu H)^\dag(\D_\mu H)-\ff{1}{\Omega^4}V(H)\Big],
\end{align}
where the fourth term is a total derivative for $\,\Omega =1\,$,
and is always suppressed by $\,1/\Mp^2\,$ for $\,\Omega \neq 1\,$ part.
We note that the form of the gauge field kinetic term does not change
under this field transformation since the Weyl factor $\,\Omega^{-4}$\,
from the determinant of the metric $\sqrt{-g\,}$\,
is compensated by additional two factors of $\,\Omega^{2}$\,
from the inverse metrics in the gauge kinetic term
$\,-\fr{1}{4}g^{\mu\nu}g^{\rh\si}F_{j\mu\rh}^aF_{j\nu\si}^a$\,.\,

\vspace*{1.5mm}

Next, we study the longitudinal weak boson scattering and the Goldstone boson scattering.
We explicitly demonstrate the longitudinal-Goldstone boson equivalence theorem
in the presence of nonminimal coupling $\,\xi\,$.\,
For this purpose, we will derive relevant scalar and gauge couplings from
the action (\ref{S-Einstein}).
We parameterize the Higgs doublet as usual,
$\,H = \big(\pi^+,\, \fr{1}{\sqrt{2}\,}(v+h^0+\ii\pi^0)\big)^T
\,$.\,
Thus, we can expand,
$\,\Omega^2 = 1+\xi\Mp^{-2}\big[2vh^0 + (h^0)^2 + 2\pi^+\pi^-+(\pi^0)^2\big]\,$.\,
From now on, we will also set Einstein frame metric to be flat,
$\,g_{\mu\nu}^{}=\eta_{\mu\nu}^{}$\,,\, since the graviton contributions from
$\,g_{\mu\nu}^{}\,$ are suppressed by more powers of $\,1/\Mp^2\,$ and thus will not
affect our following analysis of leading $\xi$ terms at $\,\order{\Mp^{-2}}\,$.\,
Then, we derive kinetic terms for the Higgs field $\,h^0\,$ and would-be Goldstone fields
$\,(\pi^\pm,\,\pi^0)\,$,
\beq
\label{eq:Hpi-kin}
  \ld_{\text{kin}}^s =\, \dis
  \ff{1}{2}\!\(\!1\!+\!\ff{6\xi^2v^2}{\Mp^2}\)\!(\pd_\mu h^0)^2\!
  +\ff{1}{2}(\pd_\mu\pi^0)^2\!
  +\pd_\mu\pi^-\pd^\mu\pi^+ .~~
\eeq
Hence, the Higgs field $\,h^0\,$ should be renormalized via,
$\,h^0\to\zeta h^0\,$,\, with $\,\zeta \equiv (1+6\xi^2v^2/\Mp^2)^{-1/2}$\,.\,
In consequence \cite{foot-1},
the Higgs boson mass is given by,
$\,m_h^2=2\lam v^2\zeta^2$,\,
and all the SM Higgs couplings should be rescaled accordingly.
Furthermore, under the Weyl transformation the original nonminimal term in (\ref{S-Jordan})
has led to new Higgs couplings in Einstein frame,
as shown in Eq.\,(\ref{S-Einstein}).
The first operator in the second line of Eq.\,(\ref{S-Einstein}) is nontrivial
when expanded to $\,\order{\xi^2/\Mp^2}\,$.\,
This and the other two derivative operators
in the second and third lines of Eq.\,(\ref{S-Einstein}) will give
$\,\xi^2 E^2/\Mp^2$\, type of contributions to the scattering processes.
Thus, they will become significant as the scattering energy $\,E\,$ increases.
Hence, for a full treatment of non-minimal coupling in Einstein frame we will
consistently take into account all these interactions.
Analyzing the second and third lines of (\ref{S-Einstein}), we systematically present
the $\xi$-dependent couplings up to the first order in $\,\Mp^{-2}$.\,
For scalar derivative interactions, we deduce the Lagrangian
$\,\Delta\ld^{ss}_{\text{int}}\,$,
\begin{align}
\label{EF-Lscalar}
& \hspace*{-1mm}
-\ff{\xi}{2\Mp^2}\!\big[|\pd_\mu\vec{\pi}|^2\!+\zeta^2(\pd_\mu h^0)^2\big]
  \big[|\vec{\pi}|^2\!+\zeta^2{h^0}^2\!\!+2v\zeta h^0\big]
\\
& \hspace*{-1mm}
-\ff{3\xi^2}{4\Mp^2}\!\big[|\vec{\pi}|^2\! +\zeta^2{h^0}^2\!\!+2v\zeta h^0\big]
   \pd^2\big[|\vec{\pi}|^2\!+\zeta^2 {h^0}^2\!\!+2v\zeta h^0\big], ~~~~~~
\n
\end{align}
where $\,|\vec{\pi}|^2=2\pi^+\pi^-+(\pi^0)^2$\, and
$\,|\pd_\mu^{}\vec{\pi}|^2 = 2\pd_\mu^{}\pi^+\pd^\mu\pi^-+(\pd_\mu\pi^0)^2$\,.\,
For Higgs couplings with weak gauge boson $WW/ZZ$, we derive the Lagrangian
$\,\Delta\ld^{hg}_{\text{int}}$\,,
\begin{align}
\label{EF-LhVV}
  &~\(2m_W^2 W^+_\mu W^{\mu -} \!+ m_Z^2 Z_\mu^2\)
\n
\\[1mm]
  &~\times\!\left[\(\ff{1}{v}-\ff{\xi v}{\Mp^2}\)\zeta h^0 +
    \ff{1}{2}\(\ff{1}{v^2}-\ff{5\xi}{\Mp^2}\)\zeta^2{h^0}^2\right] \!,
\hspace*{5mm}
\end{align}
which reduces to the SM couplings in the limit $\,\xi =0\,$.

\vspace*{1.5mm}

For completeness, we also discuss the fermion sector in the current formulation.
The kinetic term of a Dirac fermion $\Psi$ in the curved background can be written as,
\bge
\label{S-f}
\hspace*{-1mm}
S_{\text{f}}^{} \,= \!\int\!\!\di^4x\,\det(e_\nu^q)\bar\Psi\ga^p e^\mu_p
    \big(\ii\pd_\mu\! - \fr{1}{2}\omega_\mu{}^{mn}\si_{mn}\big)\Psi\,,~~~
\ede
where $\,e_\nu^q$\, is vierbein,  $\,\omega_\mu{}^{mn}$ denotes
spin connection, and \,$\si_{mn}^{}=\fr{\ii}{2}[\ga_m^{},\,\ga_n^{}]$\,.\,
We define this curved background in Jordan frame, which connects to
Einstein frame with a flat metric via
$\,g^{(J)}_{\mu\nu}=\Omega^{-2}\eta_{\mu\nu}^{}\,$.\,
Thus, we deduce the expressions,
%
$e_\mu^m =  \Omega^{-1}\de_\mu^m $\,
and
$\,\omega_\mu{}^{mn} = -\Omega^{-1}\(\de_\mu^m\pd^n\Omega-\de_\mu^n\pd^m\Omega\)$\,.\,
With these, the above kinetic term (\ref{S-f}) becomes,
\begin{align}
\label{eq:S-FermionE}
S_{\text{f}}^{} ~= \int\!\!\di^4x\,
\Big(\fr{1}{\,\Omega^3}\bar\Psi\ii\sla\pd\Psi
     + \fr{3}{\,\Omega^4}\bar\Psi(\ii\sla\pd\Omega)\Psi\Big)\,.
\hspace*{8mm}
\end{align}
We will comment on the implication of this fermionic part of action in Sec.\,3.

\vspace*{5mm}
\noindent
{\bf 3.\,Weak Boson Scattering from Higgs-Gravity Interactions}
\vspace*{3mm}

It is expected that combining Higgs-curvature coupling (\ref{eq:NMC})
with the SM makes the theory perturbatively non-renormalizable.
This is manifest in the Einstein frame action (\ref{S-Einstein}) via
$\xi$-dependent new higher dimensional operators involving Higgs doublet.
Hence, the high energy longitudinal weak boson scattering amplitude
will exhibit non-canceled $E^2$ behavior from these $\xi$-dependent interactions,
and thus lead to perturbative unitarity violation.
(Here $E$ is the scattering energy.)
From the derivative scalar couplings in (\ref{EF-Lscalar}), we will
show that the same $E^2$ behavior appears in the
corresponding Goldstone boson scattering amplitude.
Then, we present the {\it first demonstration} of
the longitudinal-Goldstone boson equivalence theorem (ET) \cite{He:1997zm},
in the presence of nonminimal coupling $\,\xi\,$.\,
This is highly nontrivial because the $\xi$-dependent
scalar derivative interactions (\ref{EF-Lscalar}) take very different forms
from the new Higgs-gauge boson couplings (\ref{EF-LhVV}).
We will further derive perturbative unitarity bound
on the Higgs-curvature coupling $\,\xi\,$
by studying the scattering amplitudes of both longitudinal weak bosons and
Goldstone bosons.

\vspace*{1.5mm}

Let us analyze the scalar scattering amplitudes among four electrically neutral states
$|\pi^+\pi^-\ra$, $\fr{1}{\sqrt 2}|\pi^0\pi^0\ra$, $\fr{1}{\sqrt 2}|h^0h^0\ra$,
and $|\pi^0 h^0\ra$.
At tree level, $|\pi^0 h^0\ra$ decouples from the other three states.
With systematical computation based on Lagrangian (\ref{EF-Lscalar}),
we derive the leading scattering amplitudes,
\beqa
\label{GoldAmp}
  \T[\pi^+\pi^- \!\to \pi^+\pi^-] &~=~& \FR{\,3\xi^2\!+\xi\,}{\Mp^2}(1+\cos\theta)E^2 \,,
\n\\
  \T[\pi^+\pi^- \!\to \pi^0\pi^0] &~=~& \FR{\,2(3\xi^2\!+\xi)\,}{\Mp^2}E^2 \,,
\n\\
  \T[\pi^+\pi^- \!\to h^0h^0] &~=~& \FR{\,2(3\xi^2\!+\xi)\,}{\Mp^2}E^2 \,,
\n\\
  \T[\pi^0\pi^0 \!\to \pi^0\pi^0] &~=~& \order{E^0} \,,
\\
  \T[\pi^0\pi^0 \!\to h^0h^0] &~=~& \FR{\,2(3\xi^2\!+\xi)\,}{\Mp^2}E^2 \,,
\n\\
  \T[h^0h^0 \to h^0h^0] &~=~& \order{E^0} \,,
\n\\
  \T[\pi^0 h^0 \to \pi^0 h^0] &~=~& -\FR{\,3\xi^2\!+\xi\,}{\Mp^2}(1-\cos\theta)E^2 \,,
\hspace*{10mm}
\n
\eeqa
where $\,E\,$ is the center-of-mass energy.\,

\vspace*{1.5mm}

In parallel, we have studied the
longitudinal weak boson scattering for
$|W_L^+W_L^-\ra$, $\fr{1}{\sqrt 2}|Z_L^0Z_L^0\ra$, $\fr{1}{\sqrt 2}|h^0h^0\ra$,
and $|Z_L^0 h^0\ra$.\,
We compute their scattering amplitudes in unitary gauge.
For instance, taking the process $\,W^+_LW^-_L\to Z_L^0Z_L^0\,$ and using
(\ref{EF-LhVV}), we derive the leading high energy
scattering amplitude in unitary gauge,
\begin{align}
\label{eq:WW-ZZ-ET}
& \mathcal{T}[W^+_LW^-_L \!\to Z_L^0Z_L^0] ~=~
\ff{\,8(3\xi^2\!+\xi)\,}{\Mp^2}
\ff{\,(E^2\!-2m_W^2)^2\,}{\,4(E^2\!-m_h^2)\,}~~~~
\n \\
& ~=~
\ff{\,2(3\xi^2\!+\xi)\,}{\Mp^2}E^2 + \order{E^0} \,.~~~~
\end{align}
This equals the Goldstone boson amplitude
$\,\mathcal{T}[\pi^+\pi^- \!\to \pi^0\pi^0]\,$ in (\ref{GoldAmp})
at $\,\order{E^2}\,$.\,
We have performed systematical analyses for all other longitudinal weak boson
scattering processes and find full agreements to the leading Goldstone
amplitudes (\ref{GoldAmp}).
These explicitly justify the validity of the ET
in the presence of Higgs-curvature coupling $\,\xi\,$,\, which was not studied before.
This also serves as a highly nontrivial self-consistency check of our scattering
amplitudes (\ref{GoldAmp}).

\vspace*{1.5mm}

A few comments are in order, concerning the $\xi$-dependent
leading scattering amplitudes (\ref{GoldAmp}).
We first note that the $\,\xi\,$ coupling enters our results by two ways.
One is through the Higgs field rescaling factor $\,\zeta=(1+6\xi^2v^2/\Mp^2)^{-1/2}$\,,\,
and the other arises from the Weyl factor $\,\Omega^2 = 1+\xi(2H^\dag H-v^2)/\Mp^2$\,.\,
Thus, at the first nontrivial order of $\,\Mp^{-2}\,$,\, we have both $\,\xi^2\,$ and $\xi$
contributions to the amplitudes (\ref{GoldAmp}).
For typical situations with $\,|\xi|\gg 1$\,,\, the $\xi^2$-terms dominate over $\xi$-terms,
and will thus control our final perturbative unitarity bound.
We can classify such processes into three categories. The first one is a universal
suppression factor $\,\zeta < 1\,$ for each Higgs field
in the Higgs boson production processes \cite{atkins}.
The second class of $\xi^2$-dependent processes are the longitudinal weak boson scattering.
We find that the {\it anomalous} quartic scalar couplings and cubic Higgs-gauge couplings
of (\ref{EF-Lscalar})-(\ref{EF-LhVV}) cause
non-canceled $\,\xi^2\,$ (and $\xi$) dependent
$E^2$-contributions in the longitudinal and Goldstone boson scattering amplitudes
[cf.\ (\ref{GoldAmp})-(\ref{eq:WW-ZZ-ET})],
which can become significant for large scattering energy $\,E$\,.\,
Hence, {\it the longitudinal $\,W_L^{}W_L^{}\,$ scattering
can provide a sensitive probe of $\,\xi\,$ coupling
via energy-enhanced leading contributions of $\,\order{\xi^2E^2/\Mp^2}\,$.\,}
The third class of $\xi^2$-involved processes are those containing the cubic Higgs coupling.
As shown in (\ref{EF-Lscalar}), such processes will also be enhanced at high energies
by the $\xi^2$-dependent derivative cubic Higgs couplings.
The high luminosity runs at LHC\,(14\,TeV), the future TeV linear colliders (ILC and CLIC)
and high energy circular $pp$ colliders (TLEP and SPPC)
will further probe such anomalous cubic Higgs couplings.
Finally, the $\,\order{\xi/\Mp^2}$\, couplings arise from the Weyl factor $\,\Omega\,$.\,
This includes bosonic and fermionic couplings of the Higgs boson,
as shown in (\ref{EF-Lscalar})-(\ref{EF-LhVV}) and (\ref{eq:S-FermionE}).
But they are negligible relative to the leading couplings of $\,\order{\xi^2/\Mp^2}$\,
for $\,\xi\gg 1\,$.\,
We also note that the graviton-exchanges in Einstein frame will contribute
to the scattering amplitudes at $\,\order{E^2/\Mp^2}\,$,\, but they are
$\xi$-independent and negligible as compared to the leading $\,\order{\xi^2E^2/\Mp^2}$\,
terms for $\,\xi\gg 1\,$.\,

\vspace*{1.5mm}

From Eq.\,(\ref{GoldAmp}), we compute the partial wave scattering amplitude
for the Goldstone bosons and Higgs boson,
\bge
a_\ell^{}(E) \,=\,
\FR{1}{32\pi}\int_{-1}^1\!\di\cos\theta\,P_\ell^{}(\cos\theta)\mathcal{T}(E,\theta)\,.
\ede
In our case, the partial wave amplitudes form a $4\times 4$ matrix
among the four states $\,|\pi^+\pi^-\ra$,\, $\fr{1}{\sqrt 2}|\pi^0\pi^0\ra$,\,
$\fr{1}{\sqrt 2}|h^0h^0\ra$,\, and $\,|\pi^0 h^0\ra$\,.\,
Thus, for $\,\ell =0$\,,\, we derive the following matrix of leading $s$-wave amplitudes,
\beqa
  a_0^{}(E)= \FR{\,(3\xi^2\!+\xi)E^2}{16\pi\Mp^2}\!
  \begin{pmatrix}
    1 & \sqrt2 & \sqrt2 & 0 \\[1mm]
    \sqrt2 & 0 & 1 & 0 \\[1mm]
    \sqrt2 & 1 & 0 & 0 \\[1mm]
    0 & 0 & 0 & -1
  \end{pmatrix} \!,~~~~~
\label{a0-4x4}
\eeqa
whose eigenvalues are,
\beqa
a_0^{\text{diag}}(E)=
\FR{\,(3\xi^2\!+\xi)E^2}{16\pi\Mp^2}\,
\text{diag}\(3,\,-1,\,-1,\,-1\)\,.~~~~~
\label{a0-diag}
\eeqa

 \begin{figure}[t]
 \begin{center}
 \includegraphics[width=0.48\textwidth]{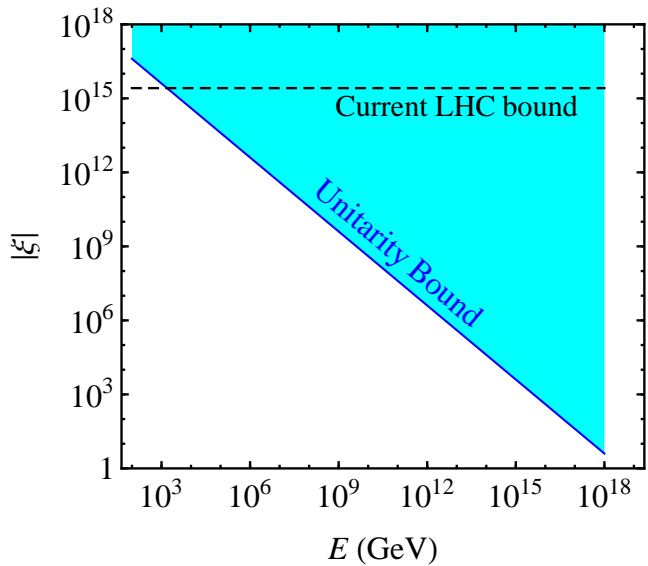}
 \vspace*{-6mm}
 \caption{Perturbative unitarity bound on Higgs-gravity coupling $\,\xi\,$ as a function of
 $WW$ scattering energy $E\,$.\, The shaded region violates perturbative unitarity.
 To compare with our bound at each given $E$ \cite{foot-3},
 the black dashed line depicts a limit  $\,|\xi|<2.7\times 10^{15}$ ($3\sigma$)
 derived from current LHC Higgs data \cite{LHC2013-3}, based on \cite{atkins}.}
 \label{fig:1}
 \end{center}
 \vspace*{-4mm}
 \end{figure}

\vspace*{1.5mm}

Then, we impose the partial wave unitarity condition,
$\,|\mathfrak{Re}\,a_0^{}| < 1/2\,$,\,
on the maximal eigenvalue of the matrix (\ref{a0-diag}).
From this, we infer the perturbative unitarity bound \cite{foot-2}
on the Higgs-curvature coupling $\,\xi\,$,
\beqa
\label{UB}
-\fr{1}{6}\big[{(C_0^2+1)^{\hf}}+1\big] \,<\, \xi \,<\,
 \fr{1}{6}\big[{(C_0^2+1)^{\hf}}-1\big] ,~~~~~~
\eeqa
where $\,C_0^{} \equiv \sqrt{32\pi\,}\Mp/E\,$.\,

%
 \begin{figure*}
 \begin{center}
 \includegraphics[width=0.47\textwidth]{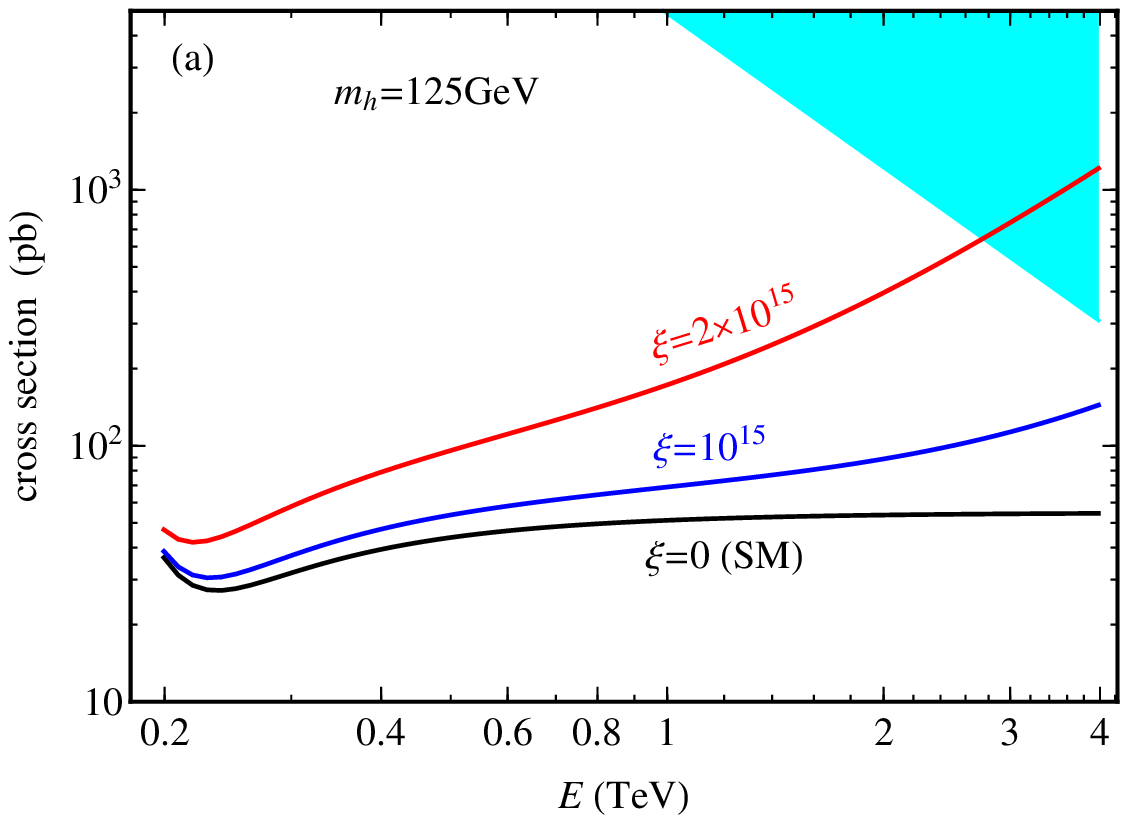}
 \hspace*{3mm}
 \includegraphics[width=0.47\textwidth]{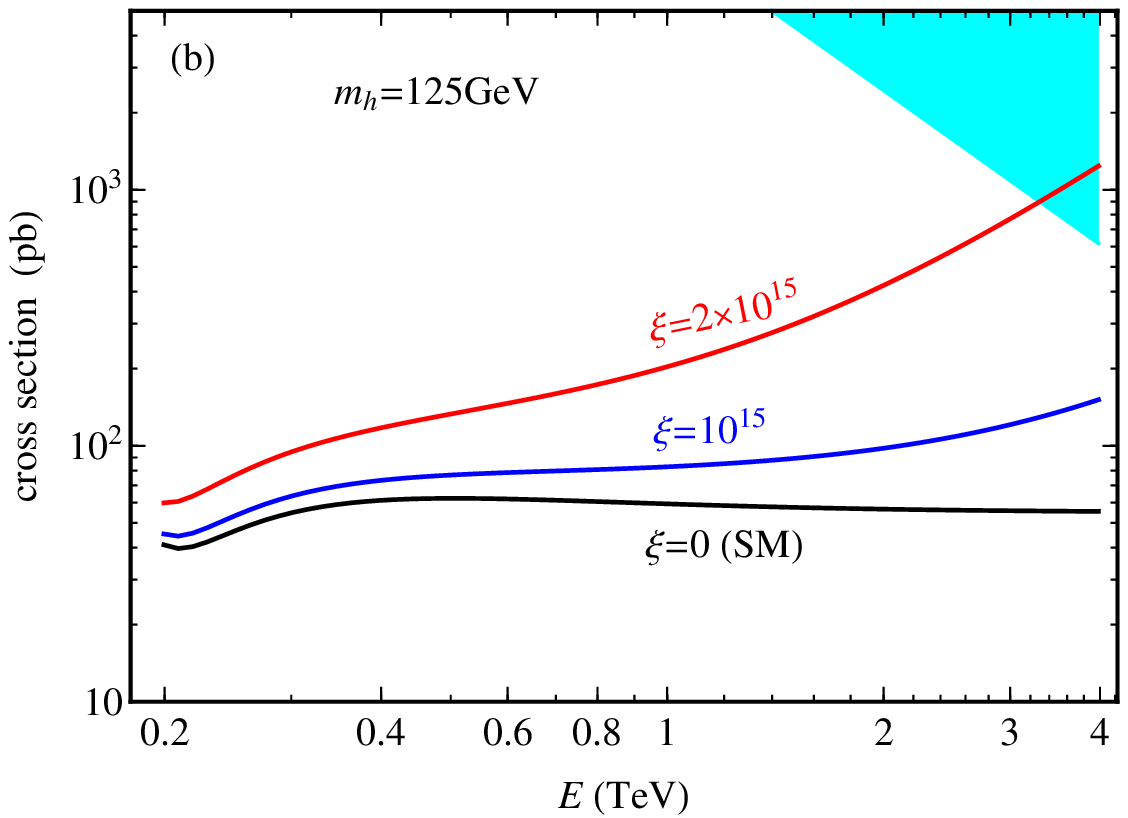}
 \vspace*{-2mm}
 \caption{Weak boson scattering cross section as a function of c.m.\ energy
 $\,E$\,.\, We present $\,W^+_L W^-_L \to Z_L^0Z_L^0\,$ in plot-(a)
 and $\,W^\pm_L W^\pm_L \to W^\pm_L W^\pm_L\,$ in plot-(b).
 In each plot, the (red,\,blue,\,black) curves depict our predictions of
 $\,\xi =(2\times\! 10^{15},\,10^{15},\,0)$,\, respectively. The black curve of
 $\,\xi=0\,$ is the pure SM expectation. We have input Higgs boson mass $\,m_h^{}=125\,$GeV.
 The shaded light-blue area shows the perturbative unitarity violation region.
 }
 \vspace*{-5mm}
 \label{fig:2}
 \end{center}
 \end{figure*}
%

\vspace*{1.5mm}

In the present effective theory formulation, we have Planck mass $\Mp$ serve as
the natural ultraviolet (UV) cutoff on the scattering energy, $\,E < \Mp\,$.\,
Thus, the inequality $\,C_0^2 > 32\pi \simeq  100.5 \gg 1\,$ holds well.
Hence, to good approximation, the above unitarity bound (\ref{UB}) reduces to
the following,
\bge
\label{UB-a}
  |\xi| ~<~ \ff{C_0^{}}{6} = \FR{\sqrt{8\pi}}{3}\FR{\Mp}{E} \,.
\ede

\vspace*{1.5mm}

In Fig.\,\ref{fig:1}, we present the perturbative unitarity bound of $\,\xi\,$ as a function
of scattering energy $E$, up to $\,E = 10^{18}\,\text{GeV}\,$,\,
which is still significantly below the Planck scale $\,\Mp\simeq 2.44\times 10^{18}\,$GeV.
We have checked both unitarity conditions (\ref{UB}) and (\ref{UB-a}), and find no
visible difference shown in the logarithmic plot of Fig.\,\ref{fig:1}.
We see that our perturbative unitarity bound
puts highly nontrivial constraints on $\,\xi\,$.\,
For the effective theory of SM\,+\,GR with Planck mass
$\,\Mp\,$ as the natural UV cutoff, the weak boson scattering energy can reach up to
$\,E = 10^{17-18}\,\text{GeV}$,\, and thus our perturbative unitarity bound requires,
$\,\xi \lesssim \order{10-1}\,$.\,
We also note that Atkins and Calmet\,\cite{atkins} recently derived
an interesting bound on $\xi$ from the 2012 LHC data \cite{LHC2012-7}.
Combining ATLAS and CMS new data in this spring \cite{LHC2013-3} gives the
Higgs signal strength, $\,\hat{\mu}=1.00\pm 0.10\,$ \cite{LHC-fit}.\,
Thus, we have a refined $3\sigma$ upper limit,  $\,|\xi|<2.7\times 10^{15}$\,.\,
For comparison, we depict this bound in Fig.\,\ref{fig:1}
by the dashed line \cite{foot-3}.
It shows that once the scattering energy $E$ exceeds
$\,\order{\text{TeV}}\,$,\, the perturbative unitarity bound becomes much more stringent.

\vspace*{1.5mm}

If Nature has chosen a much lower UV cutoff for
the effective theory of SM\,+\,GR, then a much larger $\,\xi\,$ will be allowed.
A very intriguing case is that UV cutoff lies at the TeV scale, say,
$\,\cut = \order{10\,\text{TeV}}\,$.\,
Thus, the coupling $\,\xi\,$ can reach $\,\xi = \order{10^{14-15}}\,$.\,
As a nice theory motivation for this case, it gives a conceptually simple solution to the
hierarchy problem and makes the SM Higgs sector natural up to the UV cutoff \cite{foot-4},
$\,\cut = \order{10\,\text{TeV}}\,$.\,
This further opens up an exciting possibility that
the LHC\,(14\,TeV) and its future upgrades can effectively probe
such Higgs-gravity interactions 
via weak boson scattering experiments.

\vspace*{1.5mm}

Let us study this intriguing effective theory of TeV scale quantum gravity
with $\,\cut = \order{10\,\text{TeV}}\,$.\,
It is well-known that weak boson scattering experiments serve as a key task of
LHC for probing new physics of electroweak symmetry breaking 
and Higgs mechanism \cite{WW}.
Hence, we analyze weak boson scattering cross sections, and consider two typical processes,
$\,W^+_L W^-_L \to Z_L^0Z_L^0\,$ and $\,W^\pm_L W^\pm_L \to W^\pm_L W^\pm_L\,$,\,
over the energy regime ($\,<\cut\,$) to be probed by the LHC\,(14\,TeV).
In this case, Fig.\,\ref{fig:1} shows that $\,\xi\,$
can be fairly large and reach $\,\xi = O(10^{15})\,$.\,
In Fig.\,\ref{fig:2}, we demonstrate $WW$ scattering cross sections
for two sample inputs,
$\,\xi =(2\times\! 10^{15},\,10^{15})$\,,\, as compared to
the pure SM result of $\,\xi =0\,$ \cite{foot-5}.
In addition, we impose perturbative unitarity bound on scattering cross section,
$\,\sigma < 4\pi \rho_e^{}/E^2 \,$ \cite{Dicus:2004rg},\,
as denoted by the shaded light-blue region in each plot.
(Here $\,\rho_e^{}\,$ is identical particle factor for
final state as defined in \cite{Dicus:2004rg}.)
As a remark, we clarify that there is no preferred natural values of $\,\xi\,$ from
theory side. Although the dimensionless coupling $\,\xi\,$ can have a large value,
it appears in the interaction vertices always in association with the {\it suppression factor}
$\,v^2/M_{\text{Pl}}^2$\, or $\,E^2/M_{\text{Pl}}^2$,\,
as shown in Eqs.\,\eqref{EF-Lscalar}-\eqref{EF-LhVV}.
Hence, it is fine to have a large $\,\xi\,$
as long as it holds the perturbative expansion (shown in Fig.\,\ref{fig:1}).

\vspace*{1.5mm}

From Fig.\,\ref{fig:2}(a)-(b), we see that the predicted
$WW$ scattering cross sections with sample inputs of
$\,\xi =2\times\! 10^{15},\,10^{15}$\,
exhibit different behaviors and give sizable excesses
above the pure SM expectations ($\xi =0$).
{\it Such non-resonance behaviors are universal and are expected to show up
in all $WW$ scattering channels} \cite{He:2013ub}\cite{He:2002qi},
unlike the conventional new physics models of
electroweak symmetry breaking \cite{WW}\cite{Wang:2013jwa}.
These distinctive features can be discriminated by the upcoming LHC runs at 14\,TeV
with higher luminosity.

\vspace*{7mm}
\noindent
{\bf 4.\,Conclusion and Discussion}
\vspace*{3.5mm}

The world is shaped by gravitation at the macroscopic and cosmological scales,
while the gravitational force will also play key role at the smallest Planck scale.
{\it Then, what happens in between?} With the LHC discovery of a 125\,GeV Higgs-like boson
\cite{LHC2012-7,LHC2013-3}, we are strongly motivated to explore gravitational
interactions of the Higgs boson in connection to the mechanism of electroweak symmetry
breaking and the origin of inertial mass generation for elementary particles.

\vspace*{1.9mm}

In this work, we studied Higgs-gravity interaction via the unique dimension-4
operator (\ref{eq:NMC}) with nonminimal coupling $\,\xi\,$,\,
which serves as an unambiguous portal to the new physics beyond SM.
In Sec.\,2, we focused on its formulation in Einstein frame
where this Higgs-curvature operator is transformed into
a set of $\xi$-dependent new Higgs interactions (\ref{S-Einstein}).
We derived the $\xi$-induced Higgs/Goldstone
self-interactions in Eq.\,(\ref{EF-Lscalar})
and the Higgs-gauge interactions in Eq.\,(\ref{EF-LhVV}).

\vspace*{1.9mm}

Then, in Sec.\,3 we systematically analyzed longitudinal weak boson scattering
and the corresponding Goldstone boson scattering.
We demonstrated, {\it for the first time,}
the longitudinal-Goldstone boson equivalence theorem in the presence of
Higgs-curvature coupling $\,\xi\,$.\, We performed a coupled channel analysis of
$WW$ scattering in Einstein frame and derived new perturbative unitarity bound on
$\,\xi\,$ as in Fig.\,\ref{fig:1}.
We revealed that for the SM\,+\,GR system with Planck mass
$\,\Mp\,$ as its natural UV cutoff, the weak boson scattering energy can reach
$\,E = 10^{17-18}\,\text{GeV} \,(\,< \Mp\,)$.\,
In this case, the validity of perturbative unitarity requires,
$\,\xi \lesssim \order{10-1}\,$.\,
We further studied the intriguing scenario for the SM\,+\,GR
effective theory with UV cutoff around $\,\cut = \order{10\,\text{TeV}}\,$.\,
Thus, the $\,\xi\,$ coupling can reach $\,\xi =\order{10^{15}}\,$.\,
This provides a conceptually simple resolution to the hierarchy problem
and makes the SM Higgs sector natural up to $\,\order{10\,\text{TeV}}\,$.\,
In Fig.\,\ref{fig:2}, we predicted, {\it for the first time,}
$\,WW\,$ scattering cross sections with such $\,\xi\,$ couplings, over the TeV scale.
These exhibit {\it different behaviors} from the pure SM result ($\xi =0$),
and they will be discriminated at the LHC\,(14\,TeV) with
higher integrated luminosity.
The future TeV linear colliders (ILC and CLIC)
and the future high energy circular $pp$ colliders (TLEP and SPPC)
will further probe the $\,\xi\,$ coupling via $\,WW\,$ scattering experiments.

\vspace*{1.9mm}

As a final remark, we note that
Ref.\,\cite{atkins2} studied linearized gravity
in Jordan frame, and calculated scattering amplitudes of graviton exchange
for external particles being spin-(0,\,1/2,\,1).\, It considered a nonminimal coupling term
$\,\xi R\phi^2$,\, with scalar $\phi$ having no vacuum expectation value.
It obtained a unitarity bound in the form\,\cite{atkins2},
$\,|\xi| < \order{\Mp/E}\,$,\, which has similar structure to our (\ref{UB})-(\ref{UB-a}).
But our independent analysis for the realistic SM Higgs doublet
in Einstein frame is highly nontrivial, where the nonminimal
$\xi$-term (\ref{eq:NMC}) is transformed away and the resultant $\xi$-dependent new
interactions (\ref{EF-Lscalar})-(\ref{EF-LhVV}) do not explicitly invoke gravitons.
Thus, we can compute the longitudinal/Goldstone scattering amplitudes more easily,
and extract $\xi$-dependent terms in a straightforward way. Another advantage is
that Einstein frame has canonically normalized graviton field, and the tree-level Lagrangian
manifestly preserves equivalence principle. Einstein frame has also been widely used,
including various models of Higgs inflation \cite{bezrukov,ext}.
We note that the unitarity issue of nonminimal coupling was discussed in
Einstein frame for the purpose of Higgs inflation models before\,\cite{ext}, but
those discussions are mainly power counting arguments or qualitative estimates
with rather different focus and context.
Our current work presents systematical and quantitative unitarity analysis
of Higgs-curvature interaction in Einstein frame, with physical applications
to the weak boson scattering at TeV scale and in light of the exciting
LHC Higgs discovery \cite{LHC2012-7,LHC2013-3}.
Especially, we newly demonstrate that the LHC can probe $\,\xi\,$ coupling
via weak boson scattering experiments.

\vspace*{4mm}
\noindent
Acknowledgements:\
We would like to thank X.\ Calmet, M.\ S.\ Chanowitz, F.\ Sannino, D.\ Stojkovic, L.\,C.\,R.\ Wijewardhana
and J.\,F.\ Donoghue for reading this paper and for useful discussions.
We thank A.\ Farzinnia in our group for an early discussion.
This work was supported by National NSF of China (under Grants 11275101, 11135003)
and National Basic Research Program (under Grant 2010CB833000).



\begin{thebibliography}{99}
\bibitem{foot-1}
We note that in Jordan frame, the nonminimal coupling
(\ref{eq:NMC}) will induce kinematical mixing between graviton and Higgs fields.
Its diagonalization leads to exactly the same Higgs rescaling $\,h^0\to\zeta h^0$,\,
and thus the Higgs mass formula, $\,m_h^2=2\lam v^2\zeta^2$\,.
\bibitem{foot-4}
For this effective theory study, we do not concern with any detail
of the UV dynamics above $\,\cut =\order{10\,\text{TeV}}\,$.\,
There exists well-defined TeV scale quantum gravity theories on
market. An explicit example is an extra dimensional scenario with
compactification scale of $\,\order{10\,\text{TeV}}\,$
and the Kaluza-Klein modes will
reveal themselves at energies above $\,\order{10\,\text{TeV}}\,$.
\bibitem{foot-5}
We note that the renormalization of $\,\xi\,$ coupling will cause
its running via renormalization group (RG) equation \cite{RGxi}\cite{ext}.
Since the $\,\xi\,$ running is logarithmic and our Fig.\,\ref{fig:2} has
the relevant scattering energies ($0.2 < E < 4$\,TeV) vary
within only one order of magnitude for the LHC, we find that the RG running effects
are generally small. As we have explicitly verified, the logarithmic running effect
of $\,\xi\,$ is negligible in Fig.\,\ref{fig:2}.
\end{thebibliography}
\end{document}